\documentclass[amsmath,aps,showpacs,a4paper,10pt]{revtex4}

 \usepackage{epsf}
 \usepackage{graphicx}    

\begin{document}

 \newcommand{\be}[1]{\begin{equation}\label{#1}}
 \newcommand{\ee}{\end{equation}}
 \newcommand{\bea}{\begin{eqnarray}}
 \newcommand{\eea}{\end{eqnarray}}
 \def\disp{\displaystyle}

 \def\gsim{ \lower .75ex \hbox{$\sim$} \llap{\raise .27ex \hbox{$>$}} }
 \def\lsim{ \lower .75ex \hbox{$\sim$} \llap{\raise .27ex \hbox{$<$}} }


 \title{\Large \bf On the Hojman conservation quantities in FRW Cosmology}

 \author{Aizhan Myrzakul\footnote{e-mail address: a.r.myrzakul@gmail.com}}
 \affiliation{Eurasian International Center for Theoretical Physics, Eurasian National University, Astana 010008, Kazakhstan}

 \author{Ratbay Myrzakulov\footnote{e-mail address: rmyrzakulov@gmail.com
}}
 \affiliation{Eurasian International Center for Theoretical Physics, Eurasian National University, Astana 010008, Kazakhstan}


 \begin{abstract}\vspace{1cm}
 \centerline{\bf ABSTRACT}\vspace{2mm}

In the present work we  investigate the Hojman symmetry in FRW cosmology. In particular, we use  the Hojman symmetry  to find  conserved quantities  of particular cosmological models. Next, we study the Hojman symmetry in the scalar-tensor cosmology and find  corresponding exact forms of integrals of  motion. Finally, the formalism of the Hojman symmetry  is   extended  to   dynamical systems with the higher-order equations of motion. 
 
 \end{abstract}


 \maketitle


\tableofcontents
\section{Introduction}

Symmetry plays a crucial role in the modern  theoretical and mathematical physics. Using symmetry it becomes possible to select models starting from  a fundamental laws of physics and constructing exact solutions of physical models. There are several fundamental symmetries such as Lie symmetry, Noether symmetry, Hojman symmetry, and etc. While the Lie symmetry and the Noether symmetry have been known for long time and actively used to investigate physical systems, the Hojman symmetry is a new one.   The  Noether symmetry and the Lie symmetry for  various cosmological models has widely been investigated  (see \cite{1}-\cite{4}).In contrast to them,   the Hojman symmetry for cosmological models is relatively less studied \cite{5}-\cite{7}. In contrast to Noether symmetry, in the Hojman symmetry the conservation laws can be constructed without using Hamiltonian or Lagrangian functions. It can be  obtained  by
 using just the equations of motion without refering to the Lagrangian
 or the Hamiltonian. It is interesting to note that the  conserved quantities obtained by using Hojman symmetry can be  different from the ones that are obtained in  Noether symmetry approach. Motivated by these intriguing challenge the Hojman symmetry  has
recently been extensively used to study some models of  gravity and cosmology  (see Refs. \cite{8}-\cite{14}). 

In this paper we study the Hojman symmetry for some cosmological models with the matter given by the  Chaplygin gases  as well as with the van der Waals gas. We also  generalize the Hojman symmetry for the physical systems whose equations of motion are given by the third-order and the fourth-order ordinary differential equations. The main motivation to extend the Hojman symmetry to the higher-order dynamical systems was to study it, for instance, in $F(R)$ modified gravity theory in the metric formalism. In fact, the equations of motion of the $F(R)$ gravity in the metric approach  are of the fourth-order.

The layout  of the paper is  following. In Sec. II, we briefly review some main points of the Hojman symmetry. In Sec. III, we consider the Hojman symmetry for the one and two dimensional dynamical systems given by the Friedmann equations. Sec. IV is devoted to study  some cosmological models of the universe filled by Chaplygin gases and the van der Waals gas. The Hojman symmetry for the scalar-tensor gravity theory was investigated   in Sec. V.  In Sec. VI, we consider Problem B, in which using the given conserved quantity, we can recover (find) the dynamical system (that is its equations of motion). The relation between integrable systems and the Hojman symmetry we study in Sec. VII.  Next, in Sec. VIII, we extend  the Hojman symmetry for the models with the equations of motion which are given by the third-order and fourth-order ordinary differential  equations and for some other cases. The last section is devoted to Conclusions.

\section{Brief review  of  Hojman symmetry: $q$-equations}
In this section, we provide basic elements of the Hojman symmetry for the $q$-equations \cite{6}. Let $q_{i}$ be coordinates of some physical system. We assume that they satisfy the following  set of second order ordinary differential equations
 \begin{equation}\label{1}
 \ddot{q}_{i}=F_i\left(q_j,\,\dot{q}_j,\,t\right),
\end{equation}
 where $i, j=1, \ldots, N$ and a dot stands for a derivative with respect to time $t$,  $F_i\left(q_j,\,\dot{q}_j,\,t\right)$ is a "force". In this paper, Eq.(\ref{1}) we call as the $q$-equations. Let  $q_i$ and $\tilde{q}_{i}$ be solutions of the same equations (\ref{1}) (up to $\epsilon^2$
 terms)~\cite{6}. We assume that these solutions are related by   the following infinitesimal
 transformation  
 \begin{equation}\label{2}
 \tilde{q}_{i}=q_i+\epsilon X_i\left(q_j,\,\dot{q}_j,\,t\right),
 \end{equation}
 where  $X_i=X_i\left(q_j,\,\dot{q}_j,\,t\right)$ is a symmetry
 vector for Eq.(\ref{1}). It satisfies the following set of second order linear equations ~\cite{6}
\begin{equation}\label{3}
 \frac{d^2X_i}{dt^2}-\frac{\partial F_i}{\partial q_j}X_j
 -\frac{\partial F_i}{\partial\dot{q}_j}\frac{dX_j}{dt}=0\,,
 \end{equation}
 where
\begin{equation}\label{4}
 \frac{d}{dt}=\frac{\partial}{\partial t}+
 \dot{q}_i\frac{\partial}{\partial q_i}+F_i\frac{\partial}
 {\partial\dot{q}_i}\,.
\end{equation}
Eq.(\ref{3}) is called the $X$-equation. Let the "force" $F_i$ satisfy the equation
 (in some coordinate systems)
\begin{equation}\label{5}
 \frac{\partial F_i}{\partial\dot{q}_i}=0\,.
 \end{equation}
 Then the quantity
\begin{equation}\label{6}
 Q=\frac{\partial X_i}{\partial q_i}+
 \frac{\partial}{\partial\dot{q}_i}\left(\frac{dX_i}{dt}\right)
\end{equation}
 obeys the equation
\begin{equation}\label{7}
 dQ/dt=0
\end{equation}
 that is a conserved quantity for Eq.(\ref{1}).
 Note that there exists one generalization of the last three equations. Instead of the equation (\ref{6}), let the "force"  $F_i$ satisfy (in some coordinate systems) the generalized equation
\begin{equation}\label{8}
 \frac{\partial F_i}{\partial\dot{q}_i}=-\frac{d}{dt}\ln\gamma\,.
\end{equation}
 Here we assume that  $\gamma=\gamma(q_i)$ is a function of $q_i$. In this case, the quantity $Q$ takes the form 
\begin{equation}\label{9}
 Q=\frac{1}{\gamma}\frac{\partial\left(\gamma X_i\right)}
 {\partial q_i}+\frac{\partial}{\partial\dot{q}_i}
 \left(\frac{dX_i}{dt}\right)
\end{equation}
 which is again a conserved quantity for Eq.(\ref{1}). If
 $\gamma=const.$, then Eqs.(\ref{8}) and (\ref{9}) transform  to
 Eqs.(\ref{5}) and (\ref{6}) respectively.
 
 For the pedagogical reason, here we present an example from \cite{5} which is the clasical example of the presentation to see how work the Hojman symmetry. Consider the two-dimensional harmonic oscillator. Its equations of motion reads as \cite{5}
 \begin{eqnarray}
 \ddot{q}_{1}&=&F_{1},  \label{10}\\
 \ddot{q}_{2}&=&F_{2},\label{11}
\end{eqnarray}
 where
\begin{eqnarray}
 F_{1}=(\dot{q}_{2}-\omega^{2})q_{1},  \quad F_{2}=-\frac{2\dot{q}_{1}\dot{q}_{2}}{q_{1}}. \label{12}
\end{eqnarray}
 Hence we obtain
 \begin{eqnarray}
 \frac{\partial F_{1}}{\partial \dot{q}_{1}}=0, \quad \frac{\partial F_{2}}{\partial \dot{q}_{2}}=-\frac{2\dot{q}_{1}}{q_{1}} \label{13}
 \end{eqnarray}
 so that we finally have 
  \begin{eqnarray}
 \frac{\partial F_{1}}{\partial \dot{q}_{1}}+ \frac{\partial F_{2}}{\partial \dot{q}_{2}}=-\frac{2\dot{q}_{1}}{q_{1}}=-\frac{d}{dt}\ln \gamma. \label{14}
 \end{eqnarray}
 This equation gives us
  \begin{eqnarray}
 \gamma=q^{2}_{1}. \label{15}
 \end{eqnarray}
 To find the conserved quantity we need the symmetry vectors $X_{1}$ and $X_{2}$. Their equations are given by
  \begin{eqnarray}
\frac{d}{dt}\left(\frac{dX_{1}}{dt}\right)+\omega^{2}X_{1}-\dot{q}_{2}^{2}X_{1}-2q_{1}\dot{q}_{2}\frac{dX_{2}}{dt}&=&F_{1},  \label{16}\\
\frac{d}{dt}\left(\frac{dX_{2}}{dt}\right)+\frac{2\dot{q}_{2}}{q_{1}}\frac{dX_{1}}{dt}+\frac{2\dot{q}_{1}}{q_{1}}\frac{dX_{2}}{dt}-\frac{2\dot{q}_{1}\dot{q}_{2}}{q_{1}^{2}}X_{1}&=&F_{2},\label{17}
\end{eqnarray}
 The particular solution of this set is  \cite{5}
 \begin{eqnarray}
X_{1}&=&\epsilon q^{3}_{1}\dot{q}_{2},  \label{18}\\
X_{2}&=&0. \label{19}
\end{eqnarray}
Thus we get
 \begin{eqnarray}
 \frac{\partial F_{1}}{\partial \dot{q}_{1}}+\frac{\partial F_{2}}{\partial \dot{q}_{2}}=-\frac{2\dot{q}_{1}}{q_{1}}=-\frac{d}{dt}\ln q_{1}^{2}  =-\frac{d}{dt}\ln \gamma \label{20}
\end{eqnarray}
so that 
\begin{eqnarray}
 \gamma=q_{1}^{2}. \label{21}
\end{eqnarray}
The conserved quantity $Q$ has the form \cite{5}
 \begin{eqnarray}
Q=\frac{1}{q_{1}^{2}}\frac{\partial}{\partial q_{1}}[q_{1}^{2}(q^{3}_{1}\dot{q}_{2})]+\frac{1}{q_{1}^{2}}\frac{\partial}{\partial\dot{q}_{1}}[q^{2}_{1}(q_{1}^{2}\dot{q}_{1}\dot{q}_{2})]=6q^{2}_{1}\dot{q}_{2}=const.\label{22}
\end{eqnarray}

\section{Hojman conserved quantities for Friedman equations}
We consider a spatially flat Friedmann-Robertson-Walker (FRW) universe whose line element is given by
\begin{eqnarray}
 ds^{2}=-dt^{2}+a^{2}(t)(dx^{2}+dy^{2}+dz^{2}). \label{23}
 \end{eqnarray}
For the standard General Relativity (GR) the action reads as
\begin{eqnarray}
 S=\int dx^{4}\sqrt{-g}[R+L_{m}], \label{24}
 \end{eqnarray}
 where $R$ is the Ricci scalar, $L_{m}$ is the matter Lagrangian. In the FRW space-time the Friedmann equations and the continuity equation read as
\begin{eqnarray}
 3\dot{a}^{2}-\rho a^{2}&=&0, \label{25}\\
 6\ddot{a}+(3\rho+p)a&=&0, \label{26}\\
 \dot{\rho}+3\frac{\dot{a}}{a}(\rho+p)&=&0, \label{27}
 \end{eqnarray}
 where $\rho$ and $p$ are the density and pressure of the matter. This set of equations has the following integral of motion
 \begin{eqnarray} Q=a^{n}\dot{a}^{m}+\frac{1}{6}\partial^{-1}_{t}\left(3\rho+p-\frac{2m}{3^{0.5(m-1)}}a^{m+n}\rho^{0.5(m+1)}\right) \label{28}
 \end{eqnarray}
 that is $\dot{Q}=0$, where $n, m$ are  contants and   $\partial^{-1}_{t}=\int dt$. 
 The  set of equations (\ref{25})-(\ref{27})  can be rewritten  as
\begin{eqnarray}
 3H^{2}-\rho&=&0, \label{29}\\
 2\dot{H}+\rho+p&=&0, \label{30}\\
  \dot{\rho}+3H(\rho+p)&=&0,\label{31}
\end{eqnarray}
where
\begin{eqnarray}
 H=\frac{\dot{a}}{a} \label{32}
\end{eqnarray}
 is the Hubble parameter. From these 3 equations just 2 equations are indefendent. Thus for 3 unknown functions $a, \rho, p$ or $H, \rho, p$,  we have only 2 equations. Let us return to the set (\ref{25})-(\ref{27}) or that equivalent to the set (\ref{28})-(\ref{30}). Choosing 2 equations from these 3 equations we will come to the one-domensional or two-dimensional dynamical systems.  We now consider these cases.
 
 \subsection{The one-dimensional dynamical system}
  First, we consider  the one-domensional  dynamical systems. Let us return to the set (\ref{25})-(\ref{27}) or that equivalent to the set (\ref{28})-(\ref{30}). 
  \subsubsection{Variant-1}

 \textit{Example  1.} 
  Consider the following set of equations 
  \begin{eqnarray}  
 3H^{2}-\rho&=&0, \label{33}\\
 2\dot{H}+\rho+p&=&0, \label{34}
\end{eqnarray}
  Following \cite{6} we now introduce the  coordinate $q_{1}$ as
\begin{eqnarray}
 q_{1}=\ln a. \label{35}
\end{eqnarray}
Thus $\dot{q}_{1}=H$ and the equation (\ref{32}) takes the form  
\begin{eqnarray}
 \dot{q}_{1}^{2}=\frac{1}{3}\rho. \label{36}
\end{eqnarray}
The one-dimensional dynamical system is given by
\begin{eqnarray}
  \ddot{q}_{1}=F_{1}, \label{37}
\end{eqnarray}
  where
 \begin{eqnarray}
 F_{1}=-\frac{1}{2}(3\dot{q}_{1}^{2}+p). \label{38}
\end{eqnarray}

\textit{Example  2.}
 Now let us consider the following set of equations 
  \begin{eqnarray}
 3\dot{a}^{2}-\rho a^{2}&=&0, \label{39}\\
 6\ddot{a}+(3\rho+p)a&=&0, \label{40}
 \end{eqnarray}
  The  coordinate $q_{1}$ we choose as
\begin{eqnarray}
 q_{1}=a. \label{41}
\end{eqnarray}
Thus Eq.(\ref{32}) takes the form  
\begin{eqnarray}
 \dot{q}_{1}^{2}=\frac{1}{3}\rho q_{1}^{2}. \label{42}
\end{eqnarray}
Thus in this case, the one-dimensional dynamical system reads as 
\begin{eqnarray}
  \ddot{q}_{1}=F_{1}, \label{43}
\end{eqnarray}
  where
 \begin{eqnarray}
 F_{1}=-\frac{1}{6}\left(pq_{1}+\frac{9\dot{q}_{1}^{2}}{q_{1}}\right). \label{44}
\end{eqnarray}

\subsubsection{Variant-2}

 \textit{Example  1.}  We now consider the following set of equations 
  \begin{eqnarray}
 3H^{2}-\rho&=&0, \label{45}\\
  \dot{\rho}+3H(\rho+p)&=&0,\label{46}
\end{eqnarray}
  Following \cite{6} we now introduce the  coordinate $q_{2}$ as
\begin{eqnarray}
 q_{2}=\partial^{-1}_{t}\rho \label{47}
\end{eqnarray}
or
\begin{eqnarray}
 \rho=\dot{q}_{2}. \label{48}
\end{eqnarray}
Finally we get the another version of the one-dimensional dynamical system
\begin{eqnarray}
 \ddot{q}_{2}=F_{2},\label{49}
\end{eqnarray}
 where
\begin{eqnarray}
 F_{2}=-\sqrt{3\dot{q}_{2}}(\dot{q}_{2}+p). \label{50}
 \end{eqnarray}

\textit{Example  2.}
 Now let us consider the following set of equations 
  \begin{eqnarray}
 3\dot{a}^{2}-\rho a^{2}&=&0, \label{51}\\
  \dot{\rho}+3\frac{\dot{a}}{a}(\rho+p)&=&0,\label{52}  \end{eqnarray}
  The  coordinate $q_{2}$ we choose as in (\ref{46}). Then  the one-dimensional dynamical system takes the form 
\begin{eqnarray}
  \ddot{q}_{2}=F_{2}, \label{53}
\end{eqnarray}
  where
\begin{eqnarray}
 F_{2}=-\sqrt{3\dot{q}_{2}}(\dot{q}_{2}+p) \label{54}
 \end{eqnarray}
 that is same with (\ref{49}).
 
 \subsection{The two-dimensional dynamical system}
 
 Let us return to the set of equations  (\ref{28})-(\ref{30}). As we mentioned above, in this set of equations are indefendent just two. In this subsection we consider the two-dimensional dynamical systems. 
 
 \subsubsection{Variant-1}

 Here we consider the last two equations of the set (\ref{25})-(\ref{27}) that  is
  \begin{eqnarray}
 2\dot{H}+\rho+p&=&0, \label{55}\\
  \dot{\rho}+3H(\rho+p)&=&0,\label{56}
\end{eqnarray}
 Using the definitions of the coordinates (\ref{34}) and (\ref{41}) $(H=\dot{q}_{1}, \quad \rho=\dot{q}_{2})$, the last system takes the form
 \begin{eqnarray}
 \ddot{q}_{1}&=&F_{1},  \label{57}\\
 \ddot{q}_{2}&=&F_{2},\label{58}
\end{eqnarray}
 where
\begin{eqnarray}
 F_{1}=-\frac{1}{2}(\dot{q}_{2}+p),  \quad F_{2}=-3\dot{q}_{1}(\dot{q}_{2}+p). \label{59}
\end{eqnarray}
 Hence we get
\begin{eqnarray}
 \frac{\partial F_{1}}{\partial \dot{q}_{1}}=-\frac{1}{2}\frac{\partial p}{\partial\dot{q}_{1}}, \quad \frac{\partial F_{2}}{\partial \dot{q}_{2}}=-3\dot{q}_{1}\left(1+\frac{\partial p}{\partial\dot{q}_{2}}\right) \label{60}
 \end{eqnarray}
so that we have
 \begin{eqnarray}
 \frac{\partial F_{1}}{\partial \dot{q}_{1}}+\frac{\partial F_{2}}{\partial \dot{q}_{2}}=-\frac{d}{dt}\ln \gamma \label{61}
\end{eqnarray}
 or
\begin{eqnarray}
 -\frac{1}{2}\frac{\partial p}{\partial\dot{q}_{1}}-3\dot{q}_{1}\left(1+\frac{\partial p}{\partial\dot{q}_{2}}\right)=-\frac{d}{dt}\ln \gamma. \label{62}
 \end{eqnarray}
 In this step we need the equation of state that is the explicit form of the function $p=p(\rho)$. In the next section we consider some examples.


\subsubsection{Variant-2}
 Here we consider the last two equations of the set (\ref{25})-(\ref{27}) that  is
  \begin{eqnarray}
 6\ddot{a}+(3\rho+p)a&=&0, \label{63}\\
 \dot{\rho}+3\frac{\dot{a}}{a}(\rho+p)&=&0, \label{64}
 \end{eqnarray}
 
 Using the definitions of the coordinates (\ref{34}) and (\ref{41}) $(q=a, \quad \rho=\dot{q}_{2})$ the last system takes the form
 \begin{eqnarray}
 \ddot{q}_{1}&=&F_{1},  \label{65}\\
 \ddot{q}_{2}&=&F_{2},\label{66}
\end{eqnarray}
 where
\begin{eqnarray}
 F_{1}=-\frac{(3\dot{q}_{2}+p)q_{1}}{6},  \quad F_{2}=-\frac{3\dot{q}_{1}(\dot{q}_{2}+p)}{q_{1}}. \label{67}
\end{eqnarray}
 Hence we get
\begin{eqnarray}
 \frac{\partial F_{1}}{\partial \dot{q}_{1}}=-\frac{q_{1}}{6}\frac{\partial p}{\partial\dot{q}_{1}}, \quad \frac{\partial F_{2}}{\partial \dot{q}_{2}}=-\frac{3\dot{q}_{1}}{q_{1}}\left(1+\frac{\partial p}{\partial\dot{q}_{2}}\right) \label{68}
 \end{eqnarray}
so that we have
 \begin{eqnarray}
 \frac{\partial F_{1}}{\partial \dot{q}_{1}}+\frac{\partial F_{2}}{\partial \dot{q}_{2}}=-\frac{d}{dt}\ln \gamma \label{69}
\end{eqnarray}
 or
\begin{eqnarray}
 -\frac{q_{1}}{6}\frac{\partial p}{\partial\dot{q}_{1}}-\frac{3\dot{q}_{1}}{q_{1}}\left(1+\frac{\partial p}{\partial\dot{q}_{2}}\right) =-\frac{d}{dt}\ln \gamma. \label{70}
 \end{eqnarray}
 
 In this step we need the equation of state that is the explicit form of the function $p=p(\rho)$. In the next sections we consider some examples.

 \section{Hojman conserved quantities and some cosmological gas models }
 In this section, we apply the Hojman symmetry to  some cosmological gas models. We consider the Chaplygin and modified Chaplygin gas models and the van der Waals gas model.
\subsection{Chaplygin gas}
The most popular gas is the so-called Chaplygin gas whose equation of state reads as
\begin{eqnarray}
 p=\frac{A}{\rho} \label{71}
 \end{eqnarray}
 or
\begin{eqnarray}
 p=\frac{A}{3\dot{q}_{1}^{2}}. \label{72}
 \end{eqnarray}
 Then Eq.(\ref{1}) takes the form
 \begin{eqnarray}
 \ddot{q}_{1}&=&F(\dot{q}_{1}), \label{73}
\end{eqnarray}
 where
\begin{eqnarray}
 F(\dot{q}_{1})=-\frac{1}{2}\left(3\dot{q}_{1}^{2}+\frac{A}{3\dot{q}_{1}^{2}}\right). \label{74}
 \end{eqnarray}
 Hence we get
\begin{eqnarray}
 \frac{\partial F_{1}}{\partial\dot{q}_{1}}=\frac{A}{3\dot{q}_{1}^{3}}-3\dot{q}_{1}.\label{75}
\end{eqnarray}
Substituting this expression into Eq.(\ref{8}) we obtain
\begin{eqnarray}
 \frac{A}{3\dot{q}_{1}^{3}}-3\dot{q}_{1}=-\frac{d}{dt}\ln \gamma.\label{76}
 \end{eqnarray}
 We can rewrite this equation as
\begin{eqnarray}
 \frac{A}{3\dot{q}_{1}^{4}}-3=-\frac{\partial\ln \gamma(q_{1})}{\partial q_{1}}.\label{77}
 \end{eqnarray}
 For this equation we see that the left-hand side is a function of $\dot{q_{1}}$ only. At the same time the right-hand side of this equation is a function of $q_{1}$ only. It means that they must be
 equal to a same constant in order to ensure that
 Eq.(\ref{39}) always holds. Let this
 constant be $\kappa$. Now  Eq.(\ref{39})
 can be separated into two ordinary differential equations of the form
\begin{eqnarray}
 \frac{A}{3\dot{q}_{1}^{4}}-3-\kappa &=&0, \label{78}\\
 \frac{\partial\ln \gamma(q_{1})}{\partial q_{1}}+\kappa &=&0.\label{79}
 \end{eqnarray}
 Their solutions have the forms
\begin{eqnarray}
 q_{1}&=&\left[\frac{A}{3(3+\kappa)}\right]^{0.25}t+c,\label{80} \\
  \gamma &=&\gamma_{0}e^{-\kappa q_{1}}.\label{81}
 \end{eqnarray}
 Thus the scale factor becomes
\begin{eqnarray}
 a=a_{0}e^{\left[\frac{A}{3(3+\kappa)}\right]^{0.25}t}\label{82}
 \end{eqnarray}
 and the corresponding Hubble parameter is $H=\left[\frac{A}{3(3+\kappa)}\right]^{0.25}=const.$ It is the de Sitter space-time. In this case the density and pressure take the form
\begin{eqnarray}
 \rho= \left[\frac{3A}{3+\kappa}\right]^{0.5}, \quad p=A\left[\frac{A}{3(3+\kappa)}\right]^{-0.5}.\label{83}
 \end{eqnarray}
 The EoS parameter is
\begin{eqnarray}
 \omega=\frac{p}{\rho}= 1+\frac{\kappa}{3}=-1+n,\label{84}
 \end{eqnarray}
 where
 $n=(\kappa+6)/3$.
  We now find $X_{1}$, whose equation reads as
\begin{eqnarray}
 \frac{d^2X_1}{dt^2}- \frac{\partial F_{1}}{\partial\dot{q}_{1}}\frac{dX_{1}}{dt}=0 \label{85}
 \end{eqnarray}
or
\begin{eqnarray}
\dot{q}_{1}^{2}X_{q_{1}q_{1}}+2F\dot{q}_{1}X_{q\dot{q}_{1}}+FX_{q_{1}}+F_{2}X_{\dot{q}_{1}\dot{q}_{1}}-\dot{q}_{1}F_{\dot{q}_{1}}X_{q_{1}} = 0 .\label{86}
 \end{eqnarray}
To find solutions of this equation we consider some particular cases. 

i) Example 1. $X_{1}=B(\dot{q})e^{\beta q}$. Then Eq.(\ref{48}) gives
\begin{eqnarray}
\dot{q}_{1}^{2}\beta^{2}B+2\beta F\dot{q}B_{\dot{q}}+\beta FB+F_{2}B_{\dot{q}\dot{q}}-\beta\dot{q}F_{\dot{q}}B = 0 \label{87}
 \end{eqnarray}
 or
\begin{eqnarray}
\beta^{2}y^{2}B+2\beta y FB_{y}+\beta FB+F_{2}B_{yy}-\beta yF_{y}B = 0, \label{88}
 \end{eqnarray}
 where $y=\dot{q}$. Let $B=\epsilon y^{m}$. Then we get 
\begin{eqnarray}
\beta^{2}y^{4}+2\beta m y^{2} F+\beta y^{2}F+F_{2}m(m-1)-\beta y^{3}F_{y} = 0, \label{89}
\end{eqnarray}
\begin{eqnarray}
 F(y)=-\frac{1}{2}\left(3y^{2}+\frac{A}{3y^{2}}\right), F(y)=\frac{A}{3y^{3}}-3y.\label{90}
 \end{eqnarray}
Let $m=1$. Then we obtain
 \begin{eqnarray}
\beta^{2}y^{4}+3\beta  y^{2} F-\beta y^{3}F_{y} = 0. \label{91}
 \end{eqnarray}
 This equation holds if $\beta=0$. 
 
So finally the conserved quantity $Q$ is given by
\begin{eqnarray}
Q=-\kappa\epsilon  \dot{q} +F_{\dot{q}}=\frac{A}{3\dot{q}^{3}}-3\dot{q}-\kappa\epsilon  \dot{q}=const. \label{92}
\end{eqnarray}

\subsection{Modified Chaplygin gas}

In this section, our aim is to find the Hojman symmetry for the Friedmann equations with the modified Chaplygin gas (MCG). The EoS of the MCG is given by
\begin{eqnarray}
 p=\gamma+\alpha\rho+\frac{\beta}{\rho^{n}}, \label{93}
 \end{eqnarray}
 where $\gamma, \alpha, \beta$ are real constants.  Using (\ref{21}) this formula can be rewritten as
\begin{eqnarray}
 p=\gamma+3\alpha\dot{q}^{2}+\frac{\beta}{3^{n}\dot{q}^{2n}}. \label{94}
\end{eqnarray}
 Then Eq.(\ref{25})  takes the form
\begin{eqnarray}
 \ddot{q}_{1}&=&F(\dot{q}_{1}), \label{95}
\end{eqnarray}
 where
\begin{eqnarray} 
F(\dot{q}_{1})=-\frac{1}{2}\left(\gamma+3\alpha\dot{q}_{1}^{2}+\frac{\beta}{3^{n}\dot{q}_{1}^{2n}}\right). \label{96}
\end{eqnarray}
 Hence we get
\begin{eqnarray}
 \frac{\partial F_{1}}{\partial\dot{q}_{1}}=\frac{\beta n}{3^{n}\dot{q}_{1}^{2n+1}}-3\alpha\dot{q}_{1}.\label{97}
\end{eqnarray}
Substituting these expressions into Eq.(\ref{8}) we obtain
\begin{eqnarray}
 \frac{\beta n}{3^{n}\dot{q}_{1}^{2n+1}}-3\alpha\dot{q}_{1}=-\frac{d}{dt}\ln \gamma\label{98}
\end{eqnarray}
 or
\begin{eqnarray}
 \frac{\beta n}{3^{n}\dot{q}_{1}^{2(n+1)}}-3\alpha=-\frac{\partial}{\partial q_{1}}\ln \gamma.\label{99}
\end{eqnarray}
 The solution of this equation has the form
\begin{eqnarray}
 \gamma=\gamma_{0} e^{-\kappa q_{1}}.\label{100}
\end{eqnarray}
 At the same time, from the left-side of the equation (\ref{60}) it follows that
\begin{eqnarray}
 \frac{\beta n}{3^{n}\dot{q}_{1}^{2(n+1)}}-3\alpha=\kappa.\label{101}
\end{eqnarray}
 Its solution is given by
\begin{eqnarray}
 q_{1}=c+\left[\frac{n\beta}{3^{n}(3\alpha+\kappa)}\right]^{\frac{1}{2(n+1)}}t, \label{102}
\end{eqnarray}
 from which the expressions for the scale factor and the Hubble parameter are 
\begin{eqnarray}
 a=a_{0}e^{\left[\frac{n\beta}{3^{n}(3\alpha+\kappa)}\right]^{\frac{1}{2(n+1)}}t}, \label{103}
\end{eqnarray}
 
\begin{eqnarray}
 H=\left[\frac{n\beta}{3^{n}(3\alpha+\kappa)}\right]^{\frac{1}{2(n+1)}}=const. \label{104}
\end{eqnarray}
  These expressions tell us that in this case the FRW space-time turns to the  de-Sitter one. The corresponding expressions for the density and pressure take the form
\begin{eqnarray}
 \rho= 3\left[\frac{n\beta}{3^{n}(3\alpha+\kappa)}\right]^{\frac{1}{n+1}}, \quad p=\eta+3\alpha  \left[\frac{n\beta}{3^{n}(3\alpha+\kappa)}\right]^{\frac{1}{n+1}}+\beta\left[\frac{n\beta}{3^{n}(3\alpha+\kappa)}\right]^{-\frac{n}{n+1}}.\label{105}
\end{eqnarray}
 The EoS parameter is
\begin{eqnarray}
 \omega=\frac{p}{\rho}= \frac{\eta}{3}\left[\frac{n\beta}{3^{n}(3\alpha+\kappa)}\right]^{-\frac{1}{n+1}}+\alpha+3^{-(n+1)}\beta\left[\frac{n\beta}{3^{n}(3\alpha+\kappa)}\right]^{-1}. \label{106}
\end{eqnarray}
 Let us now find $X_{1}$. The solution of the equation (\ref{3}) for $X_{1}$  we look for as $X_{1}=\epsilon\dot{q}_{1}$. It means that for this particular form of $X_{1}$ the Hojman conserved quantity has the same form as (\ref{54}).

 \subsection{The van der Waals model}
 
 Our next example is the van der Waals gas model. Its EoS reads as
\begin{eqnarray}
 p=\frac{\alpha \rho}{1-\beta\rho}-\epsilon\rho^{2}. \label{107}
\end{eqnarray}
 The van der Waals model reduces to the perfect fluid case in the limit $\beta, \epsilon\rightarrow 0$, i.e. $\lim_{\beta, \epsilon\rightarrow 0} p=\alpha \rho$. Let us rewrite Eq.(69) in terms of $q_{1}$ as
\begin{eqnarray}
 p=\frac{3\alpha \dot{q}_{1}^{2}}{1-3\beta\dot{q}_{1}^{2}}-9\epsilon\dot{q}_{1}^{4}. \label{108}
\end{eqnarray}
Then the equation of motion takes the form
\begin{eqnarray}
 \ddot{q}_{1}&=&F(\dot{q}_{1}), \label{109}
\end{eqnarray}
where
\begin{eqnarray}
 F(\dot{q}_{1})=-\frac{1}{2}\left(3\dot{q}_{1}^{2}+\frac{3\alpha \dot{q}_{1}^{2}}{1-3\beta\dot{q}_{1}^{2}}-9\epsilon\dot{q}_{1}^{4}\right). \label{110}
\end{eqnarray}
 Hence we get
\begin{eqnarray}
 \frac{\partial F_{1}}{\partial\dot{q}_{1}}=
  18\epsilon\dot{q}_{1}^{3}-3\dot{q}_{1}-\frac{3\alpha \dot{q}_{1}}{1-3\beta\dot{q}_{1}^{2}}-\frac{9\alpha\beta \dot{q}_{1}^{3}}{(1-3\beta\dot{q}_{1}^{2})^{2}}.\label{111}
 \end{eqnarray}
Substituting these expressions into Eq.(\ref{8}) we obtain
\begin{eqnarray}
  18\epsilon\dot{q}_{1}^{3}-3\dot{q}_{1}-\frac{3\alpha \dot{q}_{1}}{1-3\beta\dot{q}_{1}^{2}}-\frac{9\alpha\beta \dot{q}_{1}^{3}}{(1-3\beta\dot{q}_{1}^{2})^{2}}=-\frac{d}{dt}\ln \gamma,\label{112}
\end{eqnarray}
 or
\begin{eqnarray}
  18\epsilon\dot{q}_{1}^{2}-3-\frac{3\alpha }{1-3\beta\dot{q}_{1}^{2}}-\frac{9\alpha\beta \dot{q}_{1}^{2}}{(1-3\beta\dot{q}_{1}^{2})^{2}}=-\frac{\partial}{\partial q}\ln \gamma.\label{113}
\end{eqnarray}
 According to arguments that are same as for Eq.(\ref{61}), we get 
\begin{eqnarray}
  18\epsilon\dot{q}_{1}^{2}-3-\frac{3\alpha }{1-3\beta\dot{q}_{1}^{2}}-\frac{9\alpha\beta \dot{q}_{1}^{2}}{(1-3\beta\dot{q}_{1}^{2})^{2}}-\kappa&=&0, \label{114}\\
  -\frac{\partial}{\partial q}\ln \gamma&=&\kappa.\label{115}
\end{eqnarray}
 Let $\dot{q}_{1}^{2}=y$. Then
\begin{eqnarray}
  18\epsilon y-3-\frac{3\alpha }{1-3\beta y}-\frac{9\alpha\beta y}{(1-3\beta y)^{2}}-\kappa=0. \label{116}
\end{eqnarray}
Let $\epsilon=p=0$. Then $\kappa=-3(1+\alpha)$ that gives us
\begin{eqnarray}
 \gamma=\gamma_{0} e^{-\kappa q_{1}}.\label{117}
\end{eqnarray}
 We now find $X_{1}$. Its equation has the form
\begin{eqnarray}
 \frac{d^2X_1}{dt^2}- \frac{\partial F_{1}}{\partial\dot{q}_{1}}\frac{dX_{1}}{dt}=0 \label{118}
\end{eqnarray}
 or
\begin{eqnarray}
 \frac{d}{dt}\left(\ln\frac{dX_{1}}{dt}\right)-\frac{\partial F_{1}}{\partial\dot{q}_{1}}=0.\label{119}
\end{eqnarray}
 
\section{Hojman conserved quantities in  scalar-tensor cosmology}

We consider the FRW space-time. In this case the equations of the GR with the scalar field reads as
\begin{eqnarray}
 \frac{\dot{a}^{2}}{a^{2}}&=&\frac{1}{3}\left[0.5\dot{\phi}^{2}+V(\phi)\right], \label{120}\\
 \ddot{a}&=&\frac{a}{3}\left[V(\phi)-\dot{\phi}^{2}\right],\label{121}\\
 \ddot{\phi}&=&-3\frac{\dot{a}}{a}\dot{\phi}-V^{'}(\phi). \label{122}
\end{eqnarray}
 We now introduce two coordinates as
\begin{eqnarray}
 q_{1}=\ln a, \quad q_{2}=\phi.\label{123}
\end{eqnarray}
 In terms of $q_{1}$ and $q_{2}$, Eq.(\ref{95}) takes the form
\begin{eqnarray}
 \dot{q}_{1}^{2}=\frac{1}{3}\left[0.5\dot{q}_{2}^{2}+V(q_{2})\right]. \label{124}
\end{eqnarray}
 At the same time, Eqs.(\ref{96})-(\ref{97}) form as
\begin{eqnarray}
 \ddot{q}_{1}&=&F_{1}, \label{125} \\
  \ddot{q}_{2}&=&F_{2},\label{126}
\end{eqnarray}
  where
\begin{equation}
 F_{1}=-\frac{1}{2}\dot{\phi}^{2}=-\frac{1}{2}\dot{q}_{2}^{2},\  F_{2}=-3\dot{q}_{1}\dot{q}_{2}-V^{'}(q_{2})\label{127}
\end{equation}
 with $V^{'}=\frac{\partial V}{\partial \phi}$. Hence we get
\begin{equation}
 \frac{\partial F_{1}}{\partial\dot{q}_{1}}=0, \ \frac{\partial F_{2}}{\partial\dot{q}_{2}}=-3\dot{q}_{1}. \label{128}
\end{equation}
Substituting these expressions into Eq.(\ref{8}) we obtain
\begin{eqnarray}
 \frac{\partial F_{1}}{\partial\dot{q}_{1}}+\frac{\partial F_{2}}{\partial\dot{q}_{2}}=-3\dot{q}_{1}=-\frac{d}{dt}\ln \gamma.\label{129}
\end{eqnarray}
 Thus we have
\begin{eqnarray}
 -3\dot{q}_{1}=-\dot{q}_{1}\frac{\partial\ln \gamma}{\partial q_{1}}-\dot{q}_{2}\frac{\partial\ln \gamma}{\partial q_{2}}.\label{130}
\end{eqnarray}
This equation has the following particular solution
\begin{eqnarray}
 \gamma=\gamma_{0} e^{3q_{1}}.\label{131}
\end{eqnarray}
 We now find $X_{1}$ and $X_{2}$ whose equations have the form
\begin{eqnarray}
 \frac{d^2X_1}{dt^2}+\dot{q}_{2}\frac{dX_{2}}{dt}&=&0, \label{132}\\
 \frac{d^2X_
 {2}}{dt^2}+V^{''}X_{2}+3\dot{q}_{2}\frac{dX_{1}}{dt}+3\dot{q}_{1}\frac{dX_{2}}{dt}&=&0. \label{133}
\end{eqnarray}
After some algebra we come to the following conserved quantity
 \begin{equation}
I=\dot{q}_{2}^{2}-6 \dot{q}_{1}^{2}+2V(q_{2})=const.  \label{134}
 \end{equation}
  In fact it is not difficult to verify that $\dot{I}=0$ if $q_{1}$ and $q_{2}$ satisfy Eq.(\ref{87}) and Eq.(\ref{88}). Note that the integral of motion $I$ is nothing but the equation (\ref{95}) as $I=0$.

\section{Problem B: From $Q$ to the $q$ - equation}

\subsection{The one-dimensional case}

In the previous sections we worked according to the line 
\begin{equation}\label{135}
 Eq.(1)\Longrightarrow F_{i}\Longrightarrow X_{i}\Longrightarrow Q.
\end{equation}
It means that the equation (\ref{1}) (dynamical system) is given, we must find $Q$ using the equations (\ref{6}) or (\ref{9}). It is {\bf Problem A}. In {\bf Problem A}, we know only Eq.(\ref{1}), in other words, we know only the functions $F_{i}$ (that is $F_{i}$ are given). All other quatities we must find. Now let us consider {\bf Problem B}. In {\bf Problem B}  we  work in the vice verse direction. In {\bf Problem B}, we know just $Q$ (conserved quantity) that is $Q$ is given and all other quatities are unknown. Let the equation (\ref{1}) (dynamical system) is not given (unknown), but the conserved quantity $Q$ is given. So that we must recover  the unknown (not given) "force"$F_{i}$ that is the equation (\ref{1}) (dynamical system) starting from the known (given) conserved quantity $Q$. The corresponding work line is
\begin{equation}\label{136}
  Q\Rightarrow F_{i}\Longrightarrow X_{i}\Longrightarrow Eq.(1).
\end{equation}
 But the problem is how recover $F_{i}$ that is the dynamical system, the $q$-equations (\ref{1}). It is an idea which we try study here. For simplicity, below we consider the case when $N=1$ that just the one dimensional dynamical system. In this case we can put $q_{1}=q$,  $F_{1}=F$ and $X_{1}=X$.   Let  $Q$ is given, then $F$ we can be find using the formula
 \begin{equation}\label{137}
 F=-\frac{\frac{\partial Q}{\partial t}+\dot{q}\frac{\partial Q}{\partial q}}{\frac{\partial Q}{\partial  \dot{q}}}.
\end{equation}
which follows from the equation (\ref{7}). To find the symmetry vector $X$,  we recall the equation (\ref{6}). It we rewrite as 
\begin{equation}\label{138}
 Q=LX,
\end{equation}
where
\begin{equation}\label{139}
 L=2\frac{\partial}{\partial q}+
 \frac{\partial^{2}}{\partial t\partial \dot{q}}+\dot{q}
 \frac{\partial^{2}}{\partial q\partial \dot{q}}+ F
 \frac{\partial^{2}}{\partial \dot{q}_{2}}.
\end{equation}
It gives us 
\begin{equation}\label{140}
 X=L^{-1}Q,
\end{equation}
It is the desired expression for the symmetry vector $X$. Here we  note that, in the same time, this expression of $X$ must satisfies the equation (\ref{3}).  To demonstrate our approach, we now consider some  simple examples (toy models).

\subsubsection{Example 1: $Q=h(q)e^{\frac{\dot{q}^{2}}{2\beta}}$}
 Let the conserved quantity $Q$ is given and has the form
\begin{equation}\label{141}
 Q=h(q)e^{\frac{\dot{q}^{2}}{2\beta}},
\end{equation}
 where  $h$ is a  funstion  of $q$. From (\ref{154}) we obtain
 \begin{equation}\label{142}
 F=F(q)=-\frac{\beta \partial h}{h\partial q}=-\frac{\partial \ln{h^{\beta}}}{\partial q}.
\end{equation}

 For simplicity we now assume that $X=X(\dot{q})$. Then from (\ref{157}) we find $X$ as 
 \begin{equation}\label{143}
 X=  \left(\frac{\partial^{2}}{\partial \dot{q}^{2}}\right)^{-1}\left(\frac{Q}{F}\right),
\end{equation}
 which must satisfies the equation (\ref{3}).  The dynamical system takes the form
  \begin{equation}\label{144}
 \ddot{q}+\frac{\partial \ln{h^{\beta}}}{\partial q}=0.
\end{equation}
  
  \subsubsection{Example 2:  $Q=f(\dot{q})e^{\beta q^{n}}$ }
    
 In our second example,  the conserved quantity $Q$ is again given. For example it has the form
\begin{equation}\label{145}
 Q=f(\dot{q})e^{\beta q^{n}},
\end{equation}
 where  $f$ is a  funstion  of $\dot{q}$. From (\ref{154}) we obtain
 \begin{equation}\label{146}
 F=F(q, \dot{q})=-\frac{\beta nq^{n-1}\dot{q}f}{f^{'}}.
\end{equation}
  Then from (\ref{157}) we find $X$ as 
 \begin{equation}\label{147}
 X=  L^{-1}\left(\frac{Q}{F}\right),
\end{equation}
 which must satisfies the equation (\ref{3}). 
  The dynamical system takes the form
  \begin{equation}\label{148}
 \ddot{q}+\frac{\beta nq^{n-1}\dot{q}}{f^{\prime}(\dot{q})}=0.
\end{equation}
In particular, if $n=1$ then 
\begin{equation}\label{149}
 \ddot{q}+\frac{\beta \dot{q}f}{f^{\prime}(\dot{q})}=0.
\end{equation}

 \subsubsection{Example 3:  $Q=s(t)+f(\dot{q})h(q)$ }
    
 In our second example,  the conserved quantity $Q$ is again given. For example it has the form
\begin{equation}\label{150}
 Q=Q(t, q, \dot{q})=s(t)+f(\dot{q})h(q),
\end{equation}
 where $s$, $h$ and  $f$ are   funstions  of $t$, $q$  and $\dot{q}$, respectively. Eq.(\ref{154}) gives
 \begin{equation}\label{151}
 F=F(t, q, \dot{q})=-\frac{\dot{s}+\dot{q}fh^{\prime}}{hf^{\prime}}.
\end{equation}
  Then the dynamical system takes the form
  \begin{equation}\label{152}
 \ddot{q}+\frac{\dot{s}+\dot{q}fh^{\prime}}{hf^{\prime}}=0.
\end{equation}

\subsection{The two-dimensional case}

We now consider the two-dimensional dynamical system which we write as
 \begin{eqnarray}
 \ddot{q}_{1}+F_{1}&=&0, \label{153}\\
 \ddot{q}_{2}+F_{2}&=&0. \label{154}
\end{eqnarray}
The conserved quantity satisfies the equation 
\begin{equation}\label{155}
 0=\frac{dQ}{dt}=\frac{\partial Q}{\partial t}+
 \dot{q}_1\frac{\partial Q}{\partial q_1}+
 \dot{q}_2\frac{\partial Q}{\partial q_2}+F_1\frac{\partial Q}
 {\partial\dot{q}_1}+F_2\frac{\partial Q}
 {\partial\dot{q}_2}.
\end{equation}
We now assume that, for intance, $F_1$ is given. Then $F_{2}$ has the form
\begin{equation}\label{156}
F_{2}=-\frac{\frac{\partial Q}{\partial t}+
 \dot{q}_1\frac{\partial Q}{\partial q_1}+
 \dot{q}_2\frac{\partial Q}{\partial q_2}+F_1\frac{\partial Q}
 {\partial\dot{q}_1}}{\frac{\partial Q}
 {\partial\dot{q}_2}}
 \end{equation}
 As an example, let us consider the following some generalized two-dimensional harmonic oscillator
 \begin{eqnarray}
 \ddot{q}_{1}&=&F_{1},  \label{157}\\
 \ddot{q}_{2}&=&F_{2},\label{158}
\end{eqnarray}
 where
\begin{eqnarray}
 F_{1}=(\dot{q}_{2}-\omega^{2})q_{1} \label{159}
\end{eqnarray}
and $F_{2}$ is unknown. Let the conserved quantity $Q$ is given and has the form
\begin{equation}
Q=f(q_{1})g(q_{2})h(\dot{q}_{1})z(\dot{q}_{2}).\label{160}
 \end{equation}
 We find $F_{2}$ we use Eq.(\ref{156}). We obtain\begin{equation}
F_{2}=-\frac{\dot{q}_{1}f^{\prime}z}{fz^{\prime}}-\frac{\dot{q}_{2}g^{\prime}z}{gz^{\prime}}-\frac{(\dot{q}_{2}-\omega^{2})q_{1} z}{z^{\prime}}. \label{161}
 \end{equation}
 In particular case, when
 \begin{equation}
f=6q_{1}^{2}, \quad g=h=0, \quad z=\dot{q}_{2}^{2}\label{162}
 \end{equation}
 the quantities $F_{2}$ and $Q$ take the forms given in Eq.(\ref{12}) and (\ref{22}), respectively.
\subsection{The three-dimensional case}

We now consider the three-dimensional dynamical system (harmonic oscillator) which we write as
 \begin{eqnarray}
 \ddot{q}_{1}-F_{1}&=&0, \label{163}\\
 \ddot{q}_{2}-F_{2}&=&0, \label{164}\\
  \ddot{q}_{3}-F_{3}&=&0, \label{164}
\end{eqnarray}
where  
\begin{eqnarray}
 F_{1}=(\dot{q}_{2}-\omega^{2})q_{1},  \quad F_{2}=-\frac{2\dot{q}_{1}\dot{q}_{2}}{q_{1}} \label{12}
\end{eqnarray}
and $F_{3}$ is unknown. To find $F_{3}$ we use the formula
\begin{equation}\label{167}
F_{3}=-\frac{\frac{\partial Q}{\partial t}+
 \dot{q}_1\frac{\partial Q}{\partial q_1}+
 \dot{q}_2\frac{\partial Q}{\partial q_2}+F_1\frac{\partial Q}
 {\partial\dot{q}_1}+F_2\frac{\partial Q}
 {\partial\dot{q}_2}}{\frac{\partial Q}
 {\partial\dot{q}_3}}
 \end{equation}
 As an example, let the conserved quantity $Q$ has the form
 \begin{eqnarray}
Q=6q^{2}_{1}\dot{q}_{3}=const.\label{168}
\end{eqnarray}
Then from Eq.(\ref{167}) follows 
\begin{equation}\label{169}
F_{3}=-\frac{2 \dot{q}_1 \dot{q}_3}{ q_1}.
 \end{equation}
Finally the three-dimensional dynamical system takes the form \begin{eqnarray}
 \ddot{q}_{1}&=&(\dot{q}_{2}-\omega^{2})q_{1}, \label{170}\\
 \ddot{q}_{2}&=&-\frac{2\dot{q}_{1}\dot{q}_{2}}{q_{1}}, \label{171}\\
  \ddot{q}_{3}&=&-\frac{2 \dot{q}_1 \dot{q}_3}{ q_1}. \label{172}
\end{eqnarray}

\section{Relation between the Hojman symmetry and integrable systems}

The main aim of the Hojman symmetry is to find the conserved quatities for some dynamical systems. Some of such dynamical systems are nonlinear and integrable. As well-known such integrable nonlinear dynamical systems admit the infinite number integrals of motions. It is interesting to study the relation between the Hojman conserved quatities and integrals of motion of integrable dynamical systems. This is a question that we are going to consider in this subsection. 

\subsection{The one-dimensional case}

As an example of integrable dynamical systems we consider the Painleve - I equation.  Its equation reads as
 \begin{eqnarray}
  \ddot{q}=F, \label{173}
 \end{eqnarray}
 where 
 \begin{eqnarray}
  F=6q^{2}+t. \label{174}
 \end{eqnarray}
 The equation for the symmetry vector as the form
\begin{eqnarray}
 \frac{d^2X}{dt^2}-12qX=0.\label{175}
 \end{eqnarray}
 or
 \begin{equation}\label{176}
 \left\{\left[\frac{\partial}{\partial t}+
 \dot{q}\frac{\partial}{\partial q}+(6q^{2}+t)\frac{\partial}
 {\partial\dot{q}}\right]^{2}-12q\right\}X=0.
\end{equation}
 In our case 
\begin{equation}\label{177}
 \frac{\partial F}{\partial\dot{q}}=0
 \end{equation}
 so that   the conserved quantity is given by
\begin{equation}\label{178}
 R=\frac{\partial X}{\partial q}+
 \frac{\partial}{\partial\dot{q}}\left(\frac{dX}{dt}\right).
\end{equation}
 Thus we reached to our aim which was to establish  the statement of the problem about the relation between the Hojman symmetry and integrable one-dimensional nonlinear  systems.

\subsection{The two-dimensional case}

As an example of integrable dynamical systems we consider the Painleve - I equation.  Its equation reads as
 \begin{eqnarray}
  \ddot{q}_{1}&=&F_{1}, \label{179}\\
  \ddot{q}_{2}&=&F_{2}, \label{180}
 \end{eqnarray}
 where $F_{2}$ is unknown and 
 \begin{eqnarray}
  F_{1}=6q^{2}_{1}+t. \label{181}
 \end{eqnarray}
 The unknown force $F_{2}$ we find as
 \begin{equation}\label{1182}
F_{2}=-\frac{\frac{\partial Q}{\partial t}+
 \dot{q}_1\frac{\partial Q}{\partial q_1}+
 \dot{q}_2\frac{\partial Q}{\partial q_2}+F_1\frac{\partial Q}
 {\partial\dot{q}_1}}{\frac{\partial Q}
 {\partial\dot{q}_2}}
 \end{equation}
 Let the conserved quatity $Q$ is given and has the form
  \begin{equation}
Q=s(t)+f(\dot{q}_{2})\label{183}
 \end{equation}
 Then for  $F_{2}$  we obtain formula
 \begin{equation}
F_{2}=-\frac{\dot{s}}{f^{\prime}}.\label{184}
 \end{equation}  
 The dynamical system has the form
 \begin{eqnarray}
  \ddot{q}_{1}&=&6q^{2}_{1}+t, \label{185}\\
  \ddot{q}_{2}&=&-\frac{\dot{s(t)}}{f^{\prime}(q_{2})}. \label{186}
 \end{eqnarray}
\section{Some generalizations of the Hojman conserved theorem}

In the above we have considered the Hojman symmetry for some FRW cosmological gas models. The equations of motion of these models were second order differential equations of the form (\ref{1}). However, in some cases the equations of physical systems can be given by differential equations of the order higher than the second order (see e.g. Refs. \cite{15}-\cite{36}). In this section we try to generalize the Hojman symmetry for the third-order and fourth-order differential equations.

\subsection{The third-order differential equation}
Let the equations of motion of the physical system be given as
\begin{equation}
 \dddot{q}_{\, i}=F_i\left(t, \, q_j,\,\dot{q}_j,\, \ddot{q}_j\right), \label{187}
 \end{equation}
 where $ i, j=1, \ldots, N$, the  dot stands for a derivative with respect to time $t$ and $F_i$ is the "force".  Below we consider Problem A and Problem B. 
 \subsubsection{Problem A}
We start from Problem A. Let  $q_i$ and $\tilde{q}_{\,i}$ be solutions of the same equations (\ref{97}) (up to $\epsilon^2$
 terms). As in (\ref{1}), here we again assume that these solutions are related by   the following infinitesimal
 transformation  
\begin{equation}
 \tilde{q}_{\,i}=q_i+\epsilon X_i\left(t, \, q_j,\,\dot{q}_j,\, \ddot{q}_j\right), \label{188}
 \end{equation}
 where  $X_i$ is a symmetry
 vector for Eq.~(\ref{97}). In our case $X_{i}$  satisfies the following set of third order linear equations
 \begin{equation}
 \frac{d^3X_i}{dt^3}-\frac{\partial F_i}{\partial q_j}X_j
 -\frac{\partial F_i}{\partial\dot{q}_j}\frac{dX_j}{dt}=0\,,\label{189}
\end{equation}
 where
 \begin{equation}
 \frac{d}{dt}=\frac{\partial}{\partial t}+
 \dot{q}_i\frac{\partial}{\partial q_i}+\ddot{q}_i\frac{\partial}{\partial \dot{q}_i}+F_i\frac{\partial}
 {\partial\ddot{q}_i}\,. \label{190}
 \end{equation}
 Let the "force" $F_i$ satisfy the equation
 (in some coordinate systems)
 \begin{equation}
 \frac{\partial F_i}{\partial\ddot{q}_i}=0\,. \label{191}
\end{equation}
 Then the quantity
\begin{equation}
 Q=\frac{\partial X_i}{\partial q_i}+
 \frac{\partial}{\partial\dot{q}_i}\left(\frac{dX_i}{dt}\right)+
 \frac{\partial}{\partial\ddot{q}_i}\left(\frac{d^{2}X_i}{dt^{2}}\right) \label{192}
\end{equation}
 obeys the equation $dQ/dt=0$ that is a conserved quantity for Eq.(\ref{97}).
 
\subsubsection{Problem B}
 Now let us consider Problem B. We assume that the conserved quantity is given as
 \begin{equation}
 Q=\alpha q_{1}\ddot{q}^{3}_{1} \label{193}
\end{equation}
To find $F_{1}$ we use the following formula
\begin{equation}
 F_{1}=-\frac{\frac{\partial Q}{\partial t}+
 \dot{q}_1\frac{\partial Q}{\partial q_1}+\ddot{q}_1\frac{\partial Q}{\partial \dot{q}_1}}{\frac{\partial Q}
 {\partial\ddot{q}_1}}. \label{192}
 \end{equation}
 As result we obtain
 \begin{equation}
 F_{1}=-\frac{\dot{q}_{1}\ddot{q}_{1}}{3q_{1}}. \label{193}
 \end{equation}
 Thus the third-order differential equation takes the form
  \begin{equation}
 \dddot{q}_{1}+\frac{\dot{q}_{1}\ddot{q}_{1}}{3q_{1}}=0. \label{194}
 \end{equation}
 It admits the conserved quantity given by the equation (\ref{191}).
 \subsection{The fourth-order differential equation}
 
 Our next example is the fourth order differential equations which are also the equations of motion of some  physical system. 
 
  \subsubsection{Problem A}
  
  We write the fourth order differential equations as
\begin{equation}
 \ddddot{q}_{\, i}=F_i\left(t, \, q_j,\,\dot{q}_j,\, \ddot{q}_j, \, \dddot{q}_j\right). \label{195}
\end{equation}
  Again let  $q_i$ and $\tilde{q}_{\,i}$ are solutions of this equation (up to $\epsilon^2$ terms). They  are related by   the following infinitesimal
 transformation  
 \begin{equation}
 \tilde{q}_{\,i}=q_i+\epsilon X_i\left(t, \, q_j,\,\dot{q}_j,\, \ddot{q}_j, \, \dddot{q}_j\right). \label{195a}
 \end{equation}
 We now assume that $X_{i}$  satisfies the following set of fourth order linear equations
 \begin{equation}
 \frac{d^4X_i}{dt^4}-\frac{\partial F_i}{\partial q_j}X_j
 -\frac{\partial F_i}{\partial\dot{q}_j}\frac{dX_j}{dt}=0\,, \label{193}
 \end{equation}
 where
\begin{equation}
 \frac{d}{dt}=\frac{\partial}{\partial t}+
 \dot{q}_i\frac{\partial}{\partial q_i}+\ddot{q}_i\frac{\partial}{\partial \dot{q}_i}+\dddot{q}_i\frac{\partial}{\partial \ddot{q}_i}+F_i\frac{\partial}
 {\partial\dddot{q}_i}\,. \label{194}
 \end{equation}
 Also we assume that the "force" $F_i$ satisfies the equation
  \begin{equation}
 \frac{\partial F_i}{\partial\dddot{q}_i}=0. \label{195}
 \end{equation}
 Then the conserved quantity for the Hojman symmetry has the form 
 \begin{equation}
 Q=\frac{\partial X_i}{\partial q_i}+
 \frac{\partial}{\partial\dot{q}_i}\left(\frac{dX_i}{dt}\right)+
 \frac{\partial}{\partial\ddot{q}_i}\left(\frac{d^{2}X_i}{dt^{2}}\right)+
 \frac{\partial}{\partial\dddot{q}_i}\left(\frac{d^{3}X_i}{dt^{3}}\right), \label{196}
\end{equation}
so that $dQ/dt=0$.

\subsubsection{Problem B}
 Consider Problem B. We again assume that the conserved quantity is given and,  for instance, has the form 
  \begin{equation}
 Q=\alpha q_{1}\dot{q}_{1}\dddot{q}^{2}_{1} \label{191}
\end{equation}
To find $F_{1}$ we use the following formula
\begin{equation}
 F_{1}=-\frac{\frac{\partial Q}{\partial t}+
 \dot{q}_1\frac{\partial Q}{\partial q_1}+\ddot{q}_1\frac{\partial Q}{\partial \dot{q}_1}+\dddot{q}_1\frac{\partial Q}{\partial \ddot{q}_1}}{\frac{\partial Q}
 {\partial\dddot{q}_1}}. \label{192}
 \end{equation}
 Finally  we obtain
 \begin{equation}
 F_{1}=-\frac{(\dot{q}^{2}_{1}+q_{1}\ddot{q}_{1})\dddot{q}_{1}}{2q_{1}\dot{q}_{1}}. \label{193}
 \end{equation}
 Thus the thourth-order differential equation has  the form
  \begin{equation}
 \ddddot{q}_{1}+\frac{\dot{q}_{1}\ddot{q}_{1}}{3q_{1}}=0. \label{194}
 \end{equation}
 with  the conserved quantity given by the equation (\ref{201}).

 \subsection{The Nth-order differential equations}
 
 The previous approach we can extend to the Nth-order differential  equations of the form 
\begin{equation}
 q^{i(N)}=F_i\left(t, \, q_j,\,\dot{q}_j,\, \ddot{q}_j, \, \dddot{q}_j, ..., q^{i(N-1)}\right). \label{195}
\end{equation}
 where $q^{i(N)}$ is the $Nth$ derivative of $q_{i}$ with the respect of $t$. As above,  let  $q_i$ and $\tilde{q}_{i}$ are solutions of this equation (up to $\epsilon^2$ terms). They  are related by   the following infinitesimal
 transformation  
 \begin{equation}
 \tilde{q}_{i}=q_i+\epsilon X_i\left(t, \, q_j,\,\dot{q}_j,\, \ddot{q}_j, \, \dddot{q}_j, ..., q^{i(N-1)}\right). \label{195a}
 \end{equation}
 We now assume that $X_{i}$  satisfies the following set of fourth order linear equations
 \begin{equation}
 \frac{d^{N}X_i}{dt^N}-\frac{\partial F_i}{\partial q_j}X_j
 -\frac{\partial F_i}{\partial\dot{q}_j}\frac{dX_j}{dt}=0\,, \label{193}
 \end{equation}
 where
\begin{equation}
 \frac{d}{dt}=\frac{\partial}{\partial t}+
 \dot{q}_i\frac{\partial}{\partial q_i}+\ddot{q}_i\frac{\partial}{\partial \dot{q}_i}+\dddot{q}_i\frac{\partial}{\partial \ddot{q}_i}+...+F_i\frac{\partial}
 {\partial\dddot{q}_{i(N)}}\,. \label{194}
 \end{equation}
 Also we assume that the "force" $F_i$ satisfies the equation
  \begin{equation}
 \frac{\partial F_i}{\partial q_{i(N)}}=0. \label{195}
 \end{equation}
 Then the conserved quantity for the Hojman symmetry has the form 
 \begin{equation}
 Q=\frac{\partial X_i}{\partial q_i}+
 \frac{\partial}{\partial\dot{q}_i}\left(\frac{dX_i}{dt}\right)+
 \frac{\partial}{\partial\ddot{q}_i}\left(\frac{d_{2}X_i}{dt^{2}}\right)+
 \frac{\partial}{\partial\dddot{q}_i}\left(\frac{d^{3}X_i}{dt^{3}}\right)+...+
 \frac{\partial}{\partial q_{i(N-1)}}\left(\frac{d^{N-1}X_i}{dt^{N-1}}\right). \label{196}
\end{equation}
\subsection{Nojman symmetry for the symmetry vectors equations: $X$ - equations}
 Let us return for example to the equations of the symmetry vectors (\ref{3}). Our aim here is to try construct the Nojman symmetry for these symmetry vectors equations. To do that we  rewrite the $X$ - equations (\ref{3})  in the following "canonical" form
 \begin{equation}
 \ddot{X}_{i}=G_{i}(t, q_{j}, \dot{q}_{j}, X_{j}, \dot{X}_{j}), \label{197}
 \end{equation}
 where $G_{i}$ is a "force" for $X_{i}$ and $q_{i}$ is a solutions of the equation (\ref{1}). We now assume that two solutions of the $X$ - equations (\ref{122}) are related by   the following infinitesimal
 transformation  
 \begin{equation}
 \tilde{X}_{\,i}=X_i+\epsilon Y_i\left(t, q_{j}, \dot{q}_{j}, X_{j}, \dot{X}_{j}\right),\label{198}
 \end{equation}
 where  $Y_i=Y_i\left(t, q_{j}, \dot{q}_{j}, X_{j}, \dot{X}_{j}\right)$ is a symmetry
 vector for Eq.(\ref{122}). It satisfies the following set of second order linear equations \cite{6}
\begin{equation}\label{199}
 \frac{d^2Y_i}{dt^2}-\frac{\partial G_i}{\partial X_j}Y_j
 -\frac{\partial G_i}{\partial\dot{X}_j}\frac{dY_j}{dt}=0\,,
 \end{equation}
 where
\begin{equation}\label{200}
 \frac{d}{dt}=\frac{\partial}{\partial t}+
 \dot{X}_i\frac{\partial}{\partial X_i}+G_i\frac{\partial}
 {\partial\dot{X}_i}.
\end{equation}
Let the "force" $G_i$ satisfy the equation
 (in some coordinate systems)
\begin{equation}\label{201}
 \frac{\partial G_i}{\partial\dot{X}_i}=0\,.
 \end{equation}
 Then the quantity
\begin{equation}\label{202}
 R=\frac{\partial Y_i}{\partial X_i}+
 \frac{\partial}{\partial\dot{X}_i}\left(\frac{dY_i}{dt}\right)
\end{equation}
 obeys the equation
\begin{equation}\label{203}
 dR/dt=0
\end{equation}
 that is a conserved quantity for Eq.(\ref{122}).
 Note that there exists one generalization of the last three equations. Instead of the equation (\ref{126}), let the "force"  $G_i$ satisfy (in some coordinate systems) the generalized equation
\begin{equation}\label{204}
 \frac{\partial G_i}{\partial\dot{X}_i}=-\frac{d}{dt}\ln\gamma\,.
\end{equation}
 Here we assume that  $\gamma=\gamma(X_i)$ is a function of $X_i$. In this case, the quantity $R$ takes the form 
\begin{equation}\label{205}
 R=\frac{1}{\gamma}\frac{\partial\left(\gamma Y_i\right)}
 {\partial X_i}+\frac{\partial}{\partial\dot{X}_i}
 \left(\frac{dY_i}{dt}\right)
\end{equation}
 which is again a conserved quantity for the $X$ - equation (\ref{122}). 
 
\subsection{Nojman symmetry for  $(q-X)$ - equations}
We now consider the so-called  $(q-X)$ - equation which reads as
\begin{eqnarray}
  \ddot{q}_{i}-F_{i}(t, q_{j}, \dot{q}_{j}, X_{j}, \dot{X}_{j})&=&0, \label{206}\\
 \frac{d^2X_i}{dt^2}-\frac{\partial F_i}{\partial q_j}X_j
 -\frac{\partial F_i}{\partial\dot{q}_j}\frac{dX_j}{dt}&=&0.\label{207}
 \end{eqnarray}
Let us rewrite these coupled equations as
 \begin{eqnarray}
  \ddot{q}_{i}&=&F_{i}(t, q_{j}, \dot{q}_{j}, X_{j}, \dot{X}_{j}), \label{208}\\
 \ddot{X}_{i}&=&G_{i}(t, q_{j}, \dot{q}_{j}, X_{j}, \dot{X}_{j}), \label{209}
 \end{eqnarray}
 where $F_{i}$ and $G_{i}$ are  "forces" for $q_{i}$ and $X_{i}$, respectively. We now assume that two solutions of these equations are related by   the following infinitesimal
 transformation  
 \begin{eqnarray}
 \tilde{q}_{\,i}&=&q_i+\epsilon X_i(t, q_{j}, \dot{q}_{j}, X_{j}, \dot{X}_{j}), \label{210}\\
 \tilde{X}_{\,i}&=&X_i+\epsilon Y_i(t, q_{j}, \dot{q}_{j}, X_{j}, \dot{X}_{j}).\label{211}
 \end{eqnarray}
 They  satisfy the following set of second-order linear equations ~\cite{6}
\begin{eqnarray}
 \frac{d^2X_i}{dt^2}-\frac{\partial F_i}{\partial q_j}X_j
 -\frac{\partial F_i}{\partial\dot{q}_j}\frac{dX_j}{dt}&=&0,\label{212}\\
 \frac{d^2Y_i}{dt^2}-\frac{\partial G_i}{\partial X_j}Y_j
 -\frac{\partial G_i}{\partial\dot{X}_j}\frac{dY_j}{dt}&=&0,\label{213}
 \end{eqnarray}
 where
\begin{equation}\label{214}
 \frac{d}{dt}=\frac{\partial}{\partial t}+
 \dot{q}_i\frac{\partial}{\partial q_i}+F_i\frac{\partial}
 {\partial\dot{q}_i}+
 \dot{X}_i\frac{\partial}{\partial X_i}+G_i\frac{\partial}
 {\partial\dot{X}_i}.
\end{equation}
Let the "forces" $F_{i}$ and $G_i$ satisfy the equation
 (in some coordinate systems)
\begin{equation}\label{215}
 \frac{\partial F_i}{\partial\dot{q}_i}+\frac{\partial G_i}{\partial\dot{X}_i}=0.
 \end{equation}
 In this case,  the conserved quantity is given by
\begin{equation}\label{216}
 R=\frac{\partial X_i}{\partial q_i}+
 \frac{\partial}{\partial\dot{q}_i}\left(\frac{dX_i}{dt}\right)+\frac{\partial Y_i}{\partial X_i}+
 \frac{\partial}{\partial\dot{X}_i}\left(\frac{dY_i}{dt}\right)
\end{equation}
 that is 
\begin{equation}\label{217}
 dR/dt=0.
\end{equation}
 Let now   "forces"  $F_i$ and $G_{i}$ satisfy  the following  equation
\begin{equation}\label{218}
 \frac{\partial F_i}{\partial\dot{q}_i}+\frac{\partial G_i}{\partial\dot{X}_i}=-\frac{d}{dt}\ln\gamma,
\end{equation}
 where   $\gamma=\gamma(q_{i}, X_i)$.  As we expecting, in this case    the function  $R$ takes the form 
\begin{equation}\label{219}
 R=\frac{1}{\gamma}\frac{\partial\left(\gamma X_i\right)}
 {\partial q_i}+\frac{\partial}{\partial\dot{q}_i}
 \left(\frac{dX_i}{dt}\right)+\frac{1}{\gamma}\frac{\partial\left(\gamma Y_i\right)}
 {\partial X_i}+\frac{\partial}{\partial\dot{X}_i}
 \left(\frac{dY_i}{dt}\right).
\end{equation}
It is the  conserved quantity for the $(q-X)$ -- equation (\ref{131})-(\ref{132}). Here we must to note that the set (\ref{131})-(\ref{132}) can be reformulate as the $q$-equations. In fact, let us introduce new coordinates $p_{i}$ as
$p_{i}=q_{i}$ and $p_{i}=X_{i}$, if $i=1,..., N$ and $i=N+1, ..., 2N$, respectively. Then the $(q-X)$-equations (\ref{131})-(\ref{132})  take the form
\begin{eqnarray}
  \ddot{p}_{i}&=&E_{i}(t, q_{j}, \dot{q}_{j}, X_{j}, \dot{X}_{j}), \label{220}
 \end{eqnarray}
where $E_{i}$ is equal to $F_{i}$ and $G_{i}$, if 
 $i=1,..., N$ and $i=N+1, ..., 2N$, respectively.

\section{Conclusions}
The  accelerated expansion of the universe is one of prime problems  in modern cosmology. To explain this phenomena  there exist  various candidates - cosmological models arising  from the standard  General Relativity and its different modifications such as $F(R)$ gravity, $F(T)$ gravity and so on. Such a large number of models  raised the question of choosing among them the most realistic models. In this way, symmetry plays a crucial role to choose  fundamental cosmological models. In this context,  the Hojman symmetry approach can be a useful tool to find out such models and to construct  some class of exact solutions of cosmological models if suitable Hojman vectors are identified.  
In this paper, we studied some cosmological models using   the Hojman symmetry. As the examples  we considered Chaplygin gases and the van der Waalls  models describing the universe and found the corresponding conserved quantities. 

Originally, the Hojman symmetry was proposed for the second-order dynamical systems. However,  as we mentioned in the Introduction, in some cases the equations of motion of physical systems have the order higher  than two. Regarding it, we have extended the standard Hojman symmetry formalism  to the third-order and fourth-order differential equations. In particular, we have derived Hojman conservation laws for these two cases. These are important results as unlike the difficulty of considering this second order symmetry in some gravity theories \cite{9}, now it might be possible to apply the fourth-order generalization of the Hojman symmetry to $f(R)$ modified gravity in the metric formalism since its equations of motion are also of the fourth-order.  
 However, it seems that the real role of the Hojman symmetry in cosmology is still an open question and require further investigations. Another open problem is to understand the relations between the Hojman symmetry with other symmetries like Noether and Lie symmetries.


\begin{thebibliography}{99}\bibitem{1} S. Capozziello, M. De Laurentis, R.  Myrzakulov, Int. J. Geom. Meth. Mod. Phys. {\bf  12}, 05, 1550065   (2015) 
\bibitem{2}	S. Capozziello, M. De Laurentis, R.  Myrzakulov,  Int. J. Geom. Meth. Mod. Phys. {\bf  12}, N09, 1550095   (2015) 
\bibitem{3} K. Myrzakulov, P. Tsyba, R.  Myrzakulov, \textit{Noether symmetry in F(T)  gravity with f--essence}, [arXiv:1601.07357]
\bibitem{4}A. Aslam, M.  Jamil, R.  Myrzakulov.  Phys. Scr., {\bf  88}, 025003  (2013)
\bibitem{5}
S.~Hojman, J.\ Phys.\ A: Math. Gen. {\bf 25}, L291 (1992).

\bibitem{6}
R.~M.~Santilli, {\it Foundations of Theoretical Mechanics I},
 Springer, New York (1978);
\bibitem{7}M.~Lutzky, J.\ Phys.\ A: Math. Gen. {\bf 12}, 973 (1979).

\bibitem{8}
S.~Hojman, J.\ Phys.\ A: Math. Gen. {\bf 17}, 2399 (1984).

\bibitem{9}
S.~Capozziello and M.~Roshan,
 Phys.\ Lett.\ B, {\bf 726}, 471 (2013) 

\bibitem{10}
M.~Paolella and S.~Capozziello,
 Phys.\ Lett.\ A, {\bf 379}, 1304 (2015) 
\bibitem{11} A. Paliathanasis, P.G.L. Leach. \textit{Comment on the Hojman conservation quantities in Cosmology}, [arXiv:1503.08466]
\bibitem{12} Hao Wei, Ya-Nan Zhou, Hong-Yu Li, Xiao-Bo Zou. Astrophys. Space Sci., {\bf  360},  6 (2015) 
\bibitem{13} I. A. Bizyaev, A. V. Borisov, I. S. Mamaev. SIGMA, {\bf  12}, 012  (2016) 
\bibitem{14} Hao Wei, Hong-Yu Li, Xiao-Bo Zou. Nucl. Phys. B, {\bf 903}, 132 (2016) 
\bibitem{15} A. Paliathanasis, P.G.L. Leach, S. Capozziello. Phys. Lett. B, {\bf  755}, 8-12  (2016) 
\bibitem{16} C.-Q. Geng, Md. Wali Hossain, R. Myrzakulov, M. Sami, E.N.Saridakis. Phys. Rev. D, {\bf 92}, 023522 (2015) 
\bibitem{17} R. Myrzakulov, S.D. Odintsov, L. Sebastiani. Phys. Rev. D,  {\bf 91}, 083529 (2015) 
\bibitem{18} K. Bamba, R. Myrzakulov, S.D. Odintsov, L. Sebastiani. Phys. Rev. D,  {\bf 90}, 043505 (2014) 
\bibitem{19} Md. Wali Hossain, R. Myrzakulov, M. Sami, E.N. Saridakis. Phys. Rev. D,  {\bf 90}, 023512 (2014) 
\bibitem{20} Md. Wali Hossain, R. Myrzakulov, M. Sami, E.N. Saridakis. Phys. Rev. D,  {\bf 89}, 123513 (2014) 
\bibitem{21} K. Bamba, Md. Wali Hossain, R. Myrzakulov, S. Nojiri, M. Sami. Phys. Rev. D,  {\bf 89}, 083518 (2014) 
\bibitem{22} L. Sebastiani, G. Cognola, R. Myrzakulov, S.D. Odintsov, S. Zerbini. Phys. Rev. D,  {\bf 89}, 023518 (2014) 
\bibitem{23} L. Sebastiani, D. Momeni, R. Myrzakulov, S. D. Odintsov. Phys. Rev. D {\bf 88}, 104022 (2013) 
\bibitem{24} G. Cognola, R. Myrzakulov, L. Sebastiani, S. Zerbini. Phys. Rev. D,  {\bf 88}, 024006 (2013) 
\bibitem{25} I. Brevik, R. Myrzakulov, S. Nojiri, S. D. Odintsov. Phys. Rev. D, {\bf 86}, 063007 (2012) 
\bibitem{26} K. Bamba, R. Myrzakulov, S. Nojiri, S. D. Odintsov. Phys. Rev. D,  {\bf 85}, 104036 (2012)
\bibitem{27} A. Lopez-Revelles, R. Myrzakulov, D. Saez-Gomez. Phys. Rev. D,  {\bf 85}, 103521 (2012) 
\bibitem{28} V. Dzhunushaliev, V. Folomeev, D. Singleton, R. Myrzakulov. Phys. Rev. D,  {\bf 82}, 045032 (2010) 
\bibitem{29} D. Momeni, R. Myrzakulov, E. Gudekli. Int. J. Geom. Meth. Mod. Phys., {\bf 12}, 10 (2015)
 \bibitem{30} D. Momeni, R. Myrzakulov. \textit{Noether symmetry in Horndeski Lagrangian}, [arXiv:1410.1520]
\bibitem{31} A. Aslam, M. Jamil, D. Momeni, R. Myrzakulov, M.A. Rashid, M. Raza. Astrophys. Space Sci., {\bf 348}, 533 (2013)
  \bibitem{32} A. Aslam, M. Jamil, D. Momeni, R. Myrzakulov. Can. J. Phys., {\bf 91}, 93 (2013)
  \bibitem{33} R.  Myrzakulov. \textit{Cosmological models with non-canonical scalar and fermion fields: k-essence, f-essence and g-essence}. [arXiv:1011.4337] 
  \bibitem{33} M. Jamil, D. Momeni, R. Myrzakulov. Eur. Phys. J. C, {\bf 72}, 2137 (2012)
  \bibitem{34} M. Jamil, S. Ali, D. Momeni, R. Myrzakulov. Eur. Phys. J. C, {\bf 72}, 1998 (2012)
  \bibitem{35} M. Jamil, F. M. Mahomed, D. Momeni. Phys. Lett. B, {\bf 702}, 315 (2011)
  
  
\end{thebibliography}
\end{document}